\documentclass[a4paper]{jpconf}

\usepackage{textpos}
\usepackage{braket}
\usepackage{slashed}
\usepackage{array}
\usepackage{graphicx}
\usepackage{subfig}
\usepackage{amsmath, amssymb, amsfonts}
\usepackage{mathtools}
\usepackage{mathrsfs}
  \DeclareMathAlphabet{\mathpzc}{OT1}{pzc}{m}{it}
\usepackage{fancyhdr}
\newcommand{\D}[2]{\mathrm{d}^{#1}{#2}}

\begin{document}
\title{Thermal field theory to all orders in \\
  gradient expansion}
\author{Peter Millington\textsuperscript{1} and
  Apostolos Pilaftsis\textsuperscript{2}}
\address{\textsuperscript{1}Consortium for Fundamental Physics,
  School of Mathematics and Statistics, \\
  University of Sheffield, Sheffield S3 7RH, United Kingdom \\
  \textsuperscript{2}Consortium for Fundamental Physics,
  School of Physics and Astronomy, \\
  University of Manchester, Manchester M13 9PL, United Kingdom}
\ead{p.w.millington@sheffield.ac.uk, apostolos.pilaftsis@manchester.ac.uk}

\begin{textblock}{4}(12,-9.2)
\begin{flushright}
\begin{footnotesize}
MAN/HEP/2013/04 \\
February 2013
\end{footnotesize}
\end{flushright}
\end{textblock}

\begin{abstract}
We present  a new perturbative formulation  of non-equilibrium thermal
field  theory,   based  upon  non-homogeneous   free  propagators  and
time-dependent  vertices.  The  resulting  time-dependent diagrammatic
perturbation series  are free of pinch singularities  without the need
for  quasi-particle approximation or  effective resummation  of finite
widths.   After  arriving at  a  physically  meaningful definition  of
particle number  densities, we derive master  time evolution equations
for statistical distribution functions,  which are valid to all orders
in perturbation theory and all  orders in a gradient expansion.  For a
scalar  model,  we  make  a  loopwise truncation  of  these  evolution
equations, whilst  still capturing fast transient  behaviour, which is
found  to  be  dominated  by energy-violating  processes,  leading  to
non-Markovian evolution of memory effects.
\end{abstract}

\vspace{-0.8em}

\section{Introduction}

The   description  of  out-of-equilibrium   many-body  field-theoretic
systems  is of  increasing relevance  in theoretical  and experimental
physics at the \emph{density frontier}.  Examples range from the early
Universe  to the  deconfined  phase of  QCD,  the quark-gluon  plasma,
relevant at heavy-ion colliders, such as  RHIC and the LHC; as well as
the  internal dynamics  of compact  astro-physical phenomena,  such as
neutron stars, and condensed matter systems.

In  \cite{Millington:2012pf},  the  present  authors introduce  a  new
perturbative approach to  non-equilibrium thermal quantum field theory
and an alternative framework in  which to derive master time evolution
equations  for  macroscopic  observables.   In  contrast  to  existing
semi-classical approaches based upon  the Boltzmann equation, this new
approach  allows  the  systematic  incorporation of  finite-width  and
off-shell  effects,  without  the  need  for  effective  resummations.
Furthermore, having a well-defined underlying perturbation theory that
is free of pinch singularities,  these time evolution equations may be
truncated in a loopwise sense  whilst retaining all orders of the time
behaviour. Existing  frameworks, based upon  systems of Kadanoff--Baym
equations   \cite{Kadanoff1989},  whilst   retaining  all   orders  in
perturbation  theory, often  rely upon  the truncation  of  a gradient
expansion  in   time  derivatives   in  order  to   obtain  calculable
expressions. In this case, one necessarily makes assumptions as to the
separation of various time-scales  in these systems.  In addition, one
must  generally assume  a quasi-particle  ansatz for  the form  of the
propagators  appearing in  these  gradient expansions.   On the  other
hand,   the  loopwise-truncated  evolution   equations  of   this  new
perturbative formalism are built from non-homogeneous free propagators
and  time-dependent  vertices,   which  together  encode  spatial  and
temporal   inhomogeneity   from   tree-level   without  any   of   the
aforementioned approximations.

\section{Canonical quantization}
\label{sec:canquant}

We begin by highlighting the  details of the canonical quantization of
a   scalar   field   pertinent   to  a   perturbative   treatment   of
non-equilibrium thermal field theory.

The  \emph{time-independent}  Schr\"{o}dinger-picture field  operator,
denoted by  a subscript~$\mathrm{S}$, may  be written in  the familiar
plane-wave decomposition
\begin{equation}
  \Phi_{\mathrm{S}}(\mathbf{x};\tilde{t}_i)\ =\ \int\!\!
  \frac{\D{3}{\mathbf{p}}}{(2\pi)^3}\,\frac{1}{2E(\mathbf{p})}\;
  \Big(\,a_{\mathrm{S}}(\mathbf{p};\tilde{t}_i)e^{i\mathbf{p}\cdot\mathbf{x}}
  \: +\: a_{\mathrm{S}}^{\dag}(\mathbf{p};\tilde{t}_i)
  e^{-i\mathbf{p}\cdot\mathbf{x}}\,\Big)\; ,
\end{equation}
where   $E(\mathbf{p})\:   =\:    \sqrt{\mathbf{p}^2   +   M^2}$   and
$a_{\mathrm{S}}^{\dag}(\mathbf{p};          \tilde{t}_i)$          and
$a_{\mathrm{S}}(\mathbf{p};     \tilde{t}_i)$     are    the     usual
single-particle creation and  annihilation operators.  It is essential
to  emphasize that  we  define the  Schr\"{o}dinger-, Heisenberg-  and
Interaction   (Dirac)-pictures  to   be  coincident   at   the  finite
\emph{micro}scopic boundary time~$\tilde{t}_i$, i.e.
\begin{equation}
  \Phi_{\mathrm{S}}(\mathbf{x};\tilde{t}_i)\ =\
  \Phi_{\mathrm{H}}(\tilde{t}_i,\mathbf{x};\tilde{t}_i)\ =\
  \Phi_{\mathrm{I}}(\tilde{t}_i,\mathbf{x};\tilde{t}_i)\; .
\end{equation}
It  is at  this picture-independent  boundary time  $\tilde{t}_i$ that
initial  conditions   must  be  specified.    Implicit  dependence  on
$\tilde{t}_i$ is  marked by separation from explicit  arguments with a
semi-colon.

The       \emph{time-dependent}      interaction-picture      operator
$\Phi_{\mathrm{I}}(x;\tilde{t}_i)$   is  obtained   via   the  unitary
transformation         $\Phi_{\mathrm{I}}(x;\tilde{t}_i)\:         =\:
e^{iH_{\mathrm{S}}^0(x_0\:              -\:              \tilde{t}_i)}
\Phi_{\mathrm{S}}(\mathbf{x};\tilde{t}_i)e^{-iH_{\mathrm{S}}^0(x_0\:
  -\: \tilde{t}_i)}$  , where  $H_{\mathrm{S}}^0$ is the  free-part of
the Hamiltonian in the Schr\"{o}dinger picture. This yields
\begin{equation}
  \label{eq:PhiI0}
  \Phi_{\mathrm{I}}(x;\tilde{t}_i)\ =\ \!\int\!\!
  \frac{\D{3}{\mathbf{p}}}{(2\pi)^3}\,\frac{1}{2E(\mathbf{p})}\;
  \Big(\,a_{\mathrm{I}}(\mathbf{p},0;\tilde{t}_i)e^{-iE(\mathbf{p})x_0}
  e^{i\mathbf{p}\cdot \mathbf{x}}
  \: +\: a_{\mathrm{I}}^{\dag}(\mathbf{p},0;\tilde{t}_i)e^{iE(\mathbf{p})x_0}
  e^{-i\mathbf{p}\cdot\mathbf{x}}\,\Big)\; .
\end{equation}
Notice    that     in~(\ref{eq:PhiI0})    the    \emph{time-dependent}
interaction-picture    operators   $a_{\mathrm{I}}^{\dag}(\mathbf{p},
\tilde{t};  \tilde{t}_i)$  and $a_{\mathrm{I}}(\mathbf{p},  \tilde{t};
\tilde{t}_i)$   are   evaluated   at   the   \emph{micro}scopic   time
$\tilde{t}\:  =\:  0$.   These  operators
satisfy the commutation relation
\begin{equation}
  \label{eq:momcomrel}
  \big[\, a_{\mathrm{I}}(\mathbf{p},\tilde{t};\tilde{t}_i\,),\:
  a_{\mathrm{I}}^{\dag}(\mathbf{p}',\tilde{t}';\tilde{t}_i\,)\, \big]
  \ =\ (2\pi)^32E(\mathbf{p})\delta^{(3)}(\mathbf{p}\: -\: \mathbf{p}')
  e^{-iE(\mathbf{p})(\tilde{t}\: -\: \tilde{t}')}\; ,
\end{equation}
with all other commutators vanishing, where we obtain an overall phase
$e^{-iE(\mathbf{p})(\tilde{t}\:  -\:  \tilde{t}')}$  for  $\tilde{t}\:
\neq\: \tilde{t}'$.

In quantum statistical mechanics, we are interested in the calculation
of  Ensemble  Expectation  Values  (EEVs)  of  operators  at  a  fixed
\emph{micro}scopic time  of observation $\tilde{t}_f$.   Such EEVs are
obtained   by   taking   the   trace   with   the   density   operator
$\rho(\tilde{t}_f;\tilde{t}_i)$, i.e.
\begin{equation}
  \label{eq:EEV}
  \braket{\bullet}_{t}\ =\
  \mathcal{Z}^{-1}(t)\,\mathrm{Tr}\,
  \rho(\tilde{t}_f;\tilde{t}_i)\,\bullet\;,
\end{equation}
where $\mathcal{Z}(t)\: =\: \mathrm{Tr}\,\rho(\tilde{t}_f;\tilde{t}_i)$
is   the   partition  function.    Here,   we   have  introduced   the
\emph{macro}scopic time  $t\:=\:\tilde{t}_f\:-\:\tilde{t}_i$, which is
simply the interval of \emph{micro}scopic time between the specification
of  the boundary  conditions  and the  subsequent  observation of  the
system.

Consider  the  following observable,  which  is  the  EEV of  a
two-point product of field operators:
\begin{equation}
  \label{eq:obs}
  \mathcal{O}(\mathbf{x},\mathbf{y},\tilde{t}_f;\tilde{t}_i)\ =\ 
  \mathcal{Z}^{-1}(t)\,\mathrm{Tr}\,
  \rho(\tilde{t}_f;\tilde{t}_i)
  \Phi(\tilde{t}_f,\mathbf{x};\tilde{t}_i)
  \Phi(\tilde{t}_f,\mathbf{y};\tilde{t}_i)\;.
\end{equation}
It  has  not  been necessary  to  specify  the  picture in  which  the
operators of the RHS of  (\ref{eq:obs}) are to be interpreted. This is
because   these  operators  are   evaluated  at   \emph{equal  times}.
Potential    observables   built    from   operators    evaluated   at
\emph{different  times}  are  \emph{picture-dependent}  and  therefore
\emph{unphysical}.  In  addition, the observable  $\mathcal{O}$ should
be invariant under time translation, depending only on the macroscopic
time                             $t$,                             i.e.
$\mathcal{O}(\mathbf{x},\mathbf{y},\tilde{t}_f;\tilde{t}_i)\:    \equiv
\:\mathcal{O}(\mathbf{x},\mathbf{y},\tilde{t}_f-\tilde{t}_i;0)\:
\equiv  \:\mathcal{O}(\mathbf{x},\mathbf{y},t)$.    Notice  also  that
there  are  7  independent  coordinates: the  spatial  coordinates
$\mathbf{x}$ and  $\mathbf{y}$ and the macroscopic time  $t$.  It will
later be convenient to work in terms of the central spatial coordinate
$\mathbf{X}\:=\:(\mathbf{x}+\mathbf{y})/2$   and   the  three-momentum
$\mathbf{q}$,   conjugate   to   the   relative   spatial   coordinate
$\mathbf{R}\:=\:\mathbf{x}-\mathbf{y}$.

The density  operator of a time-dependent  and spatially inhomogeneous
background will  in general  be an intractable  incoherent sum  of all
possible $n$  to $m$ multi-particle correlations,  non-diagonal in the
Fock space.  We may account for our ignorance of this density operator
by appealing to  the remaining freedom in the  commutation relation in
(\ref{eq:momcomrel}).  In particular, we define
\begin{subequations}
\begin{align}
  \braket{a_{\mathrm{I}}(\mathbf{p},\tilde{t}_f;\tilde{t}_i)
    a_{\mathrm{I}}^{\dag}(\mathbf{p}',\tilde{t}_f;\tilde{t}_i)}_{t}\ &= \
  (2\pi)^32E(\mathbf{p})\delta^{(3)}(\mathbf{p}-\mathbf{p}')\:+\:
  2E^{1/2}(\mathbf{p})E^{1/2}(\mathbf{p}')
  f(\mathbf{p},\mathbf{p}',t)\;,\\
  \braket{a_{\mathrm{I}}^{\dag}(\mathbf{p}',\tilde{t}_f;\tilde{t}_i)
    a_{\mathrm{I}}(\mathbf{p},\tilde{t}_f;\tilde{t}_i)}_{t}\ &= \
  2E^{1/2}(\mathbf{p})E^{1/2}(\mathbf{p}')
  f(\mathbf{p},\mathbf{p}',t)\;,
\end{align}
\end{subequations}
where
$f(\mathbf{p},\mathbf{p}',t)\:=\:f^*(\mathbf{p}',\mathbf{p},t)\,$.
The          \emph{statistical          distribution         function}
$f(\mathbf{p},\mathbf{p}',t)$  is  related   to  the  particle  number
density $n(\mathbf{q},\mathbf{X},t)$ via the Wigner transform
\begin{equation}
  \label{eq:nf}
  n(\mathbf{q},\mathbf{X},t)\ =\ \!\int\!\!\frac{\D{3}{\mathbf{Q}}}{(2\pi)^3}\;
  e^{i\mathbf{Q}\cdot\mathbf{X}}\,
  f(\mathbf{q}+\mathbf{Q}/2,\mathbf{q}-\mathbf{Q}/2,t)\;.
\end{equation}
Notice that  spatial homogeneity is broken by  the explicit dependence
of $f(\mathbf{p},\mathbf{p}',t)$ on the two three-momenta $\mathbf{p}$
and $\mathbf{p}'$.   In the  thermodynamic equilibrium limit,  we have
the   correspondence:   $f(\mathbf{p},\mathbf{p}',t)\:   \rightarrow\:
f_{\mathrm{eq}}(\mathbf{p},\mathbf{p}')\:                           =\:
(2\pi)^3\delta^{(3)}(\mathbf{p}-\mathbf{p}')
f_{\mathrm{B}}\big(E(\mathbf{p})\big)\;$,                         where
$f_{\mathrm{B}}(x)\:=\:(e^{\beta  x}-1)^{-1}$  is  the  Bose--Einstein
distribution  function  and   $\beta$  is  the  inverse  thermodynamic
temperature.

\section{Schwinger--Keldysh CTP formalism}
\label{sec:CTP}

We require a path-integral approach  to generating EEVs of products of
field   operators.    Such   an   approach   is    provided   by   the
Schwinger--Keldysh CTP formalism \cite{Schwinger:1960qe,  Keldysh:1964ud}.

In  order to  obtain  the  generating functional  of  EEVs, we  insert
unitary evolution operators  to the left and the  right of the density
operator in the partition  function $\mathcal{Z}$, yielding
\begin{equation}
  \label{eq:genfunc1}
  \mathcal{Z}[\rho,J_{\pm},t]\ = \ \mathrm{Tr}\,
    \Big[\bar{\mathrm{T}}
      e^{-i\!\int_{\Omega_t}\!\D{4}{x}\,J_-(x)\Phi_{\mathrm{H}}(x)}\Big]
    \rho_{\mathrm{H}}\big(\tilde{t}_f;\tilde{t}_i\big)
    \Big[\mathrm{T}
      e^{i\!\int_{\Omega_t}\!\D{4}{x}\,J_+(x)\Phi_{\mathrm{H}}(x)}\Big]\;,
\end{equation}
where      the      spacetime     hypervolume      $\Omega_t\:\simeq\:
[-t/2,\ t/2]\times\mathbb{R}^3$  is temporally bounded.

\begin{figure}
  \begin{center}
    \includegraphics{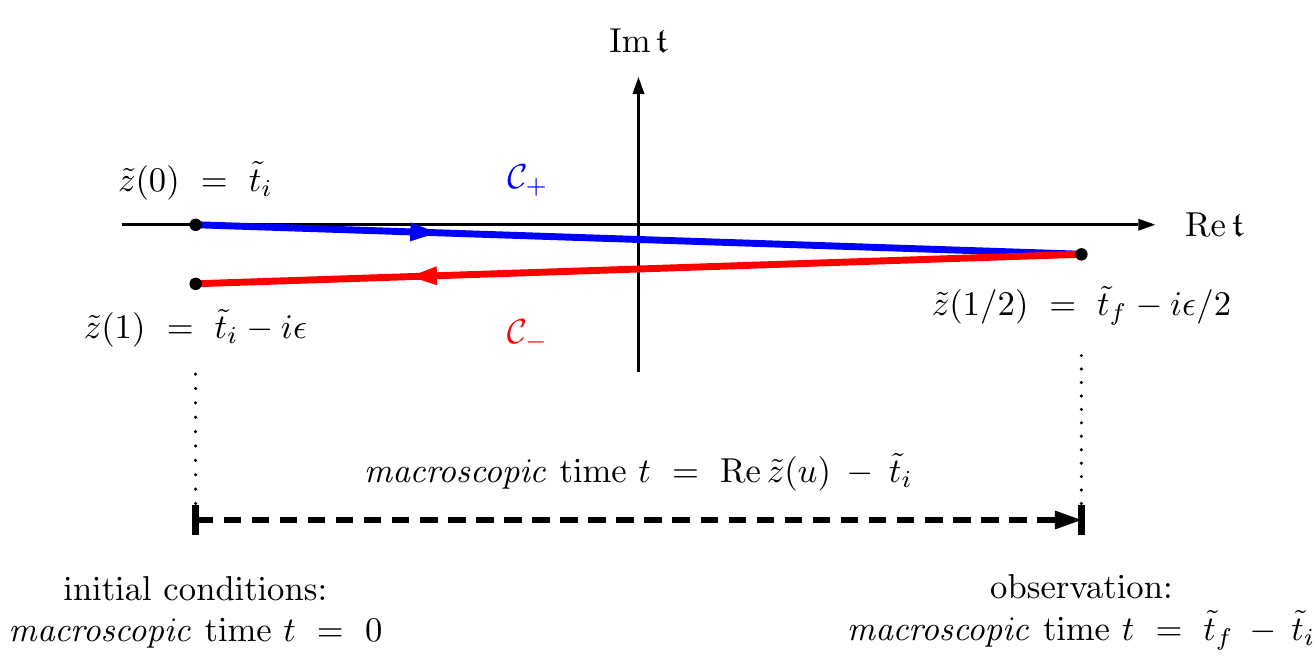}
    \vspace{-1em}
  \end{center}
  \caption{\label{fig:sk} The closed-time path,
    $\mathcal{C}\ =\ \mathcal{C}_+\:\cup\,\:\mathcal{C}_-$. The
    relationship between \emph{microscopic} and \emph{macroscopic}
    times is indicated by a dashed black arrow.}
\end{figure}

We  may in\-terpret  this  evolution as  de\-fining  a closed  contour
~$\mathcal{C}\:  =\: \mathcal{C}_+ \:  \cup\,\: \mathcal{C}_-$  in the
complex-time                plane               ($\mathfrak{t}$-plane,
$\mathfrak{t}\:\in\:\mathbb{C}$),  as  shown  in  figure~\ref{fig:sk},
which  is the  union of  two anti-parallel  branches: $\mathcal{C}_+$,
running  from   $\tilde{t}_i$  to  $\tilde{t}_f\:-\:i\epsilon/2$;  and
$\mathcal{C}_-$,  running from  $\tilde{t}_f\:-\:i\epsilon/2$  back to
$\tilde{t}_i\:-\:i\epsilon$. A small imaginary part $\epsilon\:=\:0^+$
is added to separate  the two, essentially co\-incident, branches.  We
may introduce a parametrization  of this contour $\tilde{z}(u)$, where
$u$  increases  monotonically along  $\mathcal{C}$,  which allows  the
definition  of a  path  ordering operator  $\mathrm{T}_{\mathcal{C}}$.
Notice  that,  in  contrast   to  other  interpretations  of  the  CTP
formalism,  this contour  evolves  in time,  with  each branch  having
length $t$.

Following the notation of \cite{Calzetta:1986ey, Calzetta:1986cq}, we
denote by $\Phi_{\pm}(x)\: \equiv\: \Phi(x^0 \in
\mathcal{C}_{\pm},\mathbf{x})$ fields confined to the positive and
negative branches of the CTP contour. We then define the doublets
\begin{equation}
  \Phi^a(x)\ =\ \Big(\Phi_+(x)\,,\ \Phi_-(x)\Big)\;,\qquad
  \Phi_a(x)\ =\ \eta_{ab}\Phi^b(x)\ =\ \Big(
  \Phi_+(x)\,,\ -\Phi_-(x)\Big)\; ,
\end{equation}
where    the     CTP    indices    $a,\    b\:     =\:1,\    2$    and
$\eta_{ab}\:=\:\mathrm{diag}\,(1,\  -1)$ is an  $\mathbb{SO}\,(1,\ 1)$
`metric.' 

Inserting into (\ref{eq:genfunc1}) complete sets of eigenstates of the
Heisenberg field operator, we derive a path-integral representation of
the CTP  generating functional, which depends on  the path-ordered CTP
propagator~$i\Delta^{ab}(x,y,\tilde{t}_f;\tilde{t}_i)$, written as the
$2\times 2$ matrix
\begin{equation}
  \label{eq:CTPprop}
  i\Delta^{ab}(x,y,\tilde{t}_f;\tilde{t}_i)\ \equiv\ 
  \braket{\,\mathrm{T}_{\mathcal{C}}\,
    \big[\,\Phi^a(x;\tilde{t}_i)\Phi^b(y;\tilde{t}_i)\,\big]\,}_t\ =\ 
  i\begin{bmatrix}
    \Delta_{\mathrm{F}}(x,y,\tilde{t}_f;\tilde{t}_i) &
    \Delta_{<}(x,y,\tilde{t}_f;\tilde{t}_i)
    \\ \Delta_{>}(x,y,\tilde{t}_f;\tilde{t}_i) &
    \Delta_{\mathrm{D}}(x,y,\tilde{t}_f;\tilde{t}_i)
  \end{bmatrix}\; .
\end{equation}
For          $x^0,\           y^0\in{\cal          C}_+$,          the
path-ordering~$\mathrm{T}_{\mathcal{C}}$ is equivalent to the standard
time-ordering~$\mathrm{T}$  and  we  obtain the  time-ordered  Feynman
propagator~$i\Delta_{\mathrm{F}}(x,y,\tilde{t}_f;\tilde{t}_i)$.      On
the     other     hand,      for     $x^0,\     y^0\in{\cal     C}_-$,
$\mathrm{T}_{\mathcal{C}}$          is          equivalent          to
anti-time-ordering~$\bar{\mathrm{T}}$     and     we    obtain     the
anti-time-ordered                   Dyson                  propagator~
$i\Delta_{\mathrm{D}}(x,y,\tilde{t}_f;\tilde{t}_i)$.  For $x^0\in{\cal
  C}_+$ and $y^0\in{\cal C}_-$,  $x^0$ is always `earlier' than $y^0$,
yielding    the    absolutely-ordered   negative-frequency    Wightman
propagator~$i\Delta_<(x,y,\tilde{t}_f;\tilde{t}_i)$.   Conversely, for
$y^0\in{\cal   C}_+$   and   $x^0\in{\cal   C}_-$,   we   obtain   the
positive-frequency                                             Wightman
propagator~$i\Delta_>(x,y,\tilde{t}_f;\tilde{t}_i)$.

From a Legendre transform of  the CTP generating functional, we derive
the         Cornwall--Jackiw--Tomboulis        effective        action
\cite{Cornwall:1974vz}.  We  may then obtain  the CTP Schwinger--Dyson
equation:
\begin{equation}
\label{eq:SD}
  \Delta_{ab}^{-1}(x,y,\tilde{t}_f;\tilde{t}_i)\ =\
  \Delta_{ab}^{0,-1}(x,y)\:+\:\Pi_{ab}(x,y,\tilde{t}_f;\tilde{t}_i)\;,
\end{equation}
where        $\Delta_{ab}^{-1}(x,y,\tilde{t}_f;\tilde{t}_i)$       and
$\Delta_{ab}^{0,\,-1}(x,y)$  are  the resummed  and  free inverse  CTP
propagators  and  $\Pi_{ab}(x,y,\tilde{t}_f;\tilde{t}_i)$  is the  CTP
self-energy, analogous to (\ref{eq:CTPprop}).

\section{Non-homogeneous diagrammatics}
\label{sec:nonhom}
We consider a simple scalar theory, which comprises one heavy
real scalar field  $\Phi$ and one light pair  of complex scalar fields
$(\chi^{\dag}$, $\chi)$, described by the Lagrangian
\begin{equation}
  \label{eq:model}
  \mathcal{L}\ =\ \tfrac{1}{2}\partial_{\mu}\Phi\partial^{\mu}
  \Phi\:-\:\tfrac{1}{2}M^2\Phi^2\:+\:\partial_{\mu}
  \chi^{\dag}\partial^{\mu}\chi\:-\:m^2\chi^{\dag}\chi\:-\: 
  g\Phi\chi^{\dag}\chi\:-\:\cdots\;,
\end{equation}
where  $M\:=\:1\ \mathrm{GeV}$  and  $m\:=\:0.01\ \mathrm{GeV}$.   The
ellipsis   contains  omitted   self-interactions  and   the  spacetime
dependence of the fields has been suppressed.  We prepare two isolated
but      coincident      subsystems      $\mathscr{S}_{\Phi}$      and
$\mathscr{S}_{\chi}$, both separately  in thermodynamic equilibrium at
the  same temperature $T\:=\:10\  \mathrm{GeV}$ with  the interactions
switched  off.  The subsystem  $\mathscr{S}_{\Phi}$ contains  only the
field $\Phi$ and $\mathscr{S}_{\chi}$,  only $\chi$.  At $t\:=\:0$, we
turn      on      the       interactions      and      the      system
$\mathscr{S}\:=\:\mathscr{S}_{\Phi}\cup\mathscr{S}_{\chi}$
re-thermalizes.   The subsystem  $\mathscr{S}_{\chi}$ is  taken  to be
infinite   so   that   it   is  unperturbed   by   interactions   with
$\mathscr{S}_{\Phi}$.

\begin{figure}
  \begin{center}
    \vspace{-1.5em}
    \includegraphics{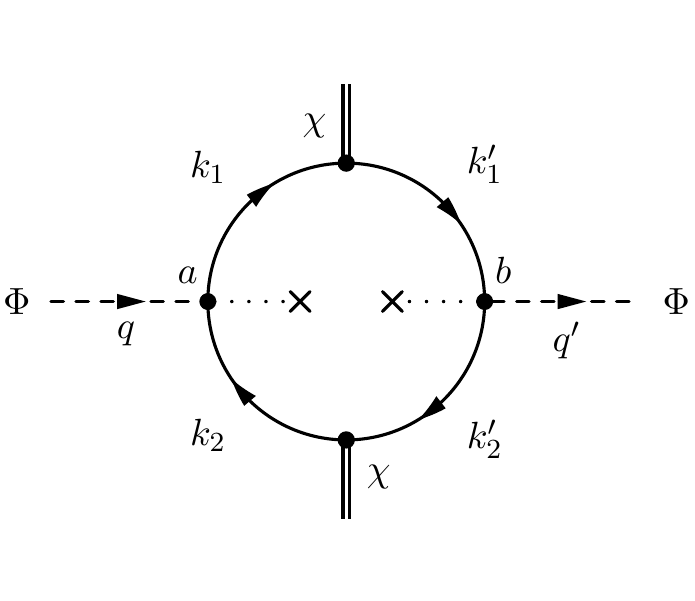}
    \vspace{-2.5em}
  \end{center}
  \caption{The non-local one-loop $\Phi$ self-energy,
    $i\Pi^{(1)}_{\Phi,\,ab}(q,q',\tilde{t}_f;\tilde{t}_i)$.}
  \label{fig:selfs}
\end{figure}

The  one-loop   non-local  $\Phi$  self-energy  is   shown  in  figure
\ref{fig:selfs}.  In particular, we  draw attention to two features of
the modified Feynman rules.   Firstly, with the vertices, we associate
a term
\begin{equation}
  \label{eq:vertex}
  -ig\,e^{iq_0\tilde{t}_f}\frac{t}{2\pi}\mathrm{sinc}
  \Big[\Big(\textstyle{\sum_i} p_{0,i}\Big)
  \frac{t}{2}\Big]\delta^{(3)}\Big(\textstyle{\sum_i}\mathbf{p}_i\Big),
\end{equation}
where the  $p_i=q,\ k_1,\ k_2$  are the four-momenta flowing  into the
vertex.  The  phase $e^{iq_0\tilde{t}_f}$,  where $q_0$ is  the energy
flow external  to the loop,  results from the proper  consideration of
the Wick contraction and field-particle duality relations.  Due to the
finite upper and  lower bounds on time integrals  appearing in the CTP
generating  functional,  the  familiar energy-conserving  Dirac  delta
function has  been replaced by  a sinc function  in (\ref{eq:vertex}).
This violation of energy conservation is shown diagrammatically by the
dotted line  terminated in  a cross and  results from  the uncertainty
principle, since the  observation of the system is  made over a finite
time interval.   Furthermore, by virtue of  this energy-violation, the
perturbation series remains free of the pinch singularities that would
otherwise  result  from products  of  delta  functions with  identical
arguments at early times.  Secondly, the double lines occurring in the
CTP propagators  of the loop  reflect the violation  of three-momentum
due to  the dependence  on the inhomogeneous  statistical distribution
function    $f(\mathbf{p},\mathbf{p}',t)$.     The    full   set    of
non-homogeneous free propagators are listed in table \ref{tab:nonhom}.
Together, these modified Feynman  rules encode the time-dependence and
spatial inhomogeneity of the  system from tree-level. All four-momenta
internal  to  the  loop  are   integrated  over  and  that  the  usual
combinatorial  factors  apply.   The  CTP indices  $a,\  b\:=\:1,\  2$
indicate the location of the vertex on either the positive or negative
branches of the CTP contour.

\begin{table}
  \begin{center}
    \begin{tabular}{m{2.4cm} m{12cm}}
      \br
      {\bfseries Propagator} & {\bfseries Double-Momentum Representation}
      \\
      \mr
      Feynman (Dyson) &
      $\begin{array}{l}
        \vspace{-0.8em} \\
        i\Delta^0_{\mathrm{F}(\mathrm{D})}(p,p',\tilde{t}_f;\tilde{t}_i)\ =\
        \displaystyle \frac{(-)i}{p^2-M^2+(-)i\epsilon}
        (2\pi)^4\delta^{(4)}(p-p')\\
        \qquad +2\pi|2p_0|^{1/2}\delta(p^2-M^2)\tilde{f}(p,p',t)
        e^{i(p_0-p_0')\tilde{t}_f}2\pi|2p_0'|^{1/2}\delta(p'^2-M^2)
      \end{array}$ \\
      $+$($-$)ve-freq. Wightman &
      $\begin{array}{l}
        \vspace{-0.5em} \\
        i\Delta_{>(<)}^0(p,p',\tilde{t}_f;\tilde{t}_i)\ =\
        2\pi\theta(+(-)p_0)\delta(p^2-M^2)(2\pi)^4\delta^{(4)}(p-p')\\
        \qquad +2\pi|2p_0|^{1/2}\delta(p^2-M^2)\tilde{f}(p,p',t)
        e^{i(p_0-p_0')\tilde{t}_f}2\pi|2p_0'|^{1/2}\delta(p'^2-M^2)
      \end{array}$\\ 
      Retarded \newline (Advanced) &
      $\begin{array}{l}
        \vspace{-0.5em} \\
        i\Delta_{\mathrm{R}(\mathrm{A})}^0(p,p')\ =\
        \displaystyle \frac{i}{(p_0+(-)i\epsilon)^2-\mathbf{p}^2-M^2}
        (2\pi)^4\delta^{(4)}(p-p')
      \end{array}$ \\
      Pauli--Jordan &
      $\begin{array}{l}
        \vspace{-0.5em} \\
        i\Delta^0(p,p')\ =\ 2\pi\varepsilon(p_0)
        \delta(p^2-M^2)(2\pi)^4\delta^{(4)}(p-p')
      \end{array}$\\
      Hadamard &
      $\begin{array}{l}
        \vspace{-0.5em} \\
        i\Delta_1^0(p,p',\tilde{t}_f;\tilde{t}_i)\ =\
        2\pi\delta(p^2-M^2)(2\pi)^4\delta^{(4)}(p-p')\\
        \qquad +2\pi|2p_0|^{1/2}\delta(p^2-M^2)2\tilde{f}(p,p',t)
        e^{i(p_0-p_0')\tilde{t}_f}2\pi|2p_0'|^{1/2}\delta(p'^2-M^2)
      \end{array}$\\
      Principal-part &
      $\begin{array}{l}
        \vspace{-0.5em} \\
        \vspace{0.4em} \displaystyle i\Delta^0_{\mathcal{P}}(p,p')\ =\
        \mathcal{P}\frac{i}{p^2-M^2}(2\pi)^4\delta^{(4)}(p-p')
      \end{array}$\\
      \br
      \end{tabular}
  \end{center}
  \vspace{-1em}
  \caption{The full complement of non-homogeneous free propagators for
    the scalar field $\Phi$, where $\tilde{f}(p,p',t)\: =\:
    \theta(p_0)\theta(p_0')f(\mathbf{p},\mathbf{p}',t)
    \:+\:\theta(-p_0)\theta(-p_0')f^*(-\mathbf{p},-\mathbf{p}',t)$.}
  \label{tab:nonhom}
\end{table}

\begin{figure}
  \begin{center}
    \subfloat[$1\rightarrow 2$ decay]{\label{fig:kinsa}
      {\centering \includegraphics[scale = 1]{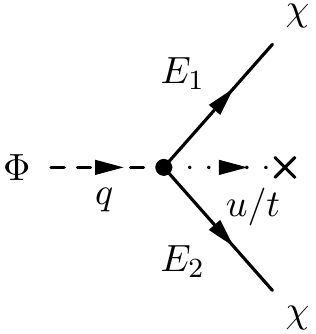}}\quad}
    \subfloat[$2\rightarrow 1$ Landau damping]{\label{fig:kinsb}
      {\centering \includegraphics[scale = 1]{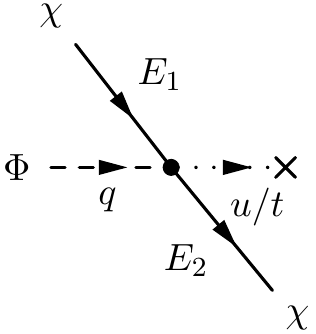} \quad
        \includegraphics[scale = 1]{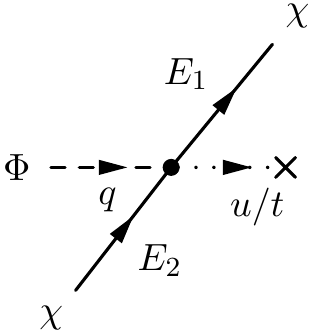}}}
    \subfloat[$3\rightarrow 0$ total annihilation]{\label{fig:kinsc}
      {\centering \qquad \includegraphics[scale = 1]{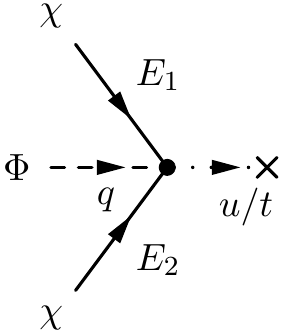} \qquad}}
    \vspace{-1em}
  \end{center}
  \caption{\label{fig:kins} The four processes contributing to the
    one-loop time-dependent $\Phi$ width.}
\vspace{-0.5em}
\end{figure}

For the system $\mathscr{S}$, the one-loop time-dependent $\Phi$ width
is given by the following integral:
\vspace{-0.5em}\begin{equation}
  \label{eq:phiwidth}
  \Gamma^{(1)}_{\Phi}(q,t)\: =\: \frac{g^2t}{64\pi^3M}\!
  \sum_{\alpha_{1},\,\alpha_{2}}\int\!\D{3}{\mathbf{k}}\;
  \frac{\alpha_1\alpha_2}{E_1E_2}\,
  \mathrm{sinc}\big[\big(q_0-\alpha_1E_1-\alpha_2E_2\big)t\big]
  \big(1+f_{\mathrm{B}}(\alpha_1E_1)+
  f_{\mathrm{B}}(\alpha_2E_2)\big)\;,
\vspace{-0.5em}\end{equation}  where $\alpha_1,\  \alpha_2\:=\:\pm 1$,
$E_1\:\equiv\:  E_{\chi}(\mathbf{k})\:=\:\sqrt{\mathbf{k}^2+m^2}$  and
$E_2\:\equiv\:E_{\chi}(\mathbf{q}-\mathbf{k})$.   The   violation   of
energy conservation, due to  the sinc function in (\ref{eq:phiwidth}),
leads     to     otherwise-forbidden     contributions    from     for
$\alpha_1,\     \alpha_2\:=\:-1$      (total     annihilation)     and
$\alpha_1\:=\:-\alpha_2$   (Landau    damping).   In   addition,   the
kinematically-allowed  phase space  for  normal $1  \to  2$ decays  is
expanded.   These   evanescent   processes   are   shown   in   figure
\ref{fig:kins}. For $t\:\to\:\infty$, we recover the known equilibrium
result, since
\vspace{-0.5em}\begin{equation}
  \label{eq:tinf}
  \lim_{t\:\to\:\infty}\,\frac{t}{\pi}\,\mathrm{sinc}\big[
  \big(q_0-\alpha_1E_1-\alpha_2
  E_2\big)t\big]\ =\
  \delta\big(q_0-\alpha_1E_1-
  \alpha_2E_2\big)\;.
\vspace{-0.5em}\end{equation}

\begin{figure}
  \begin{center}
    \includegraphics{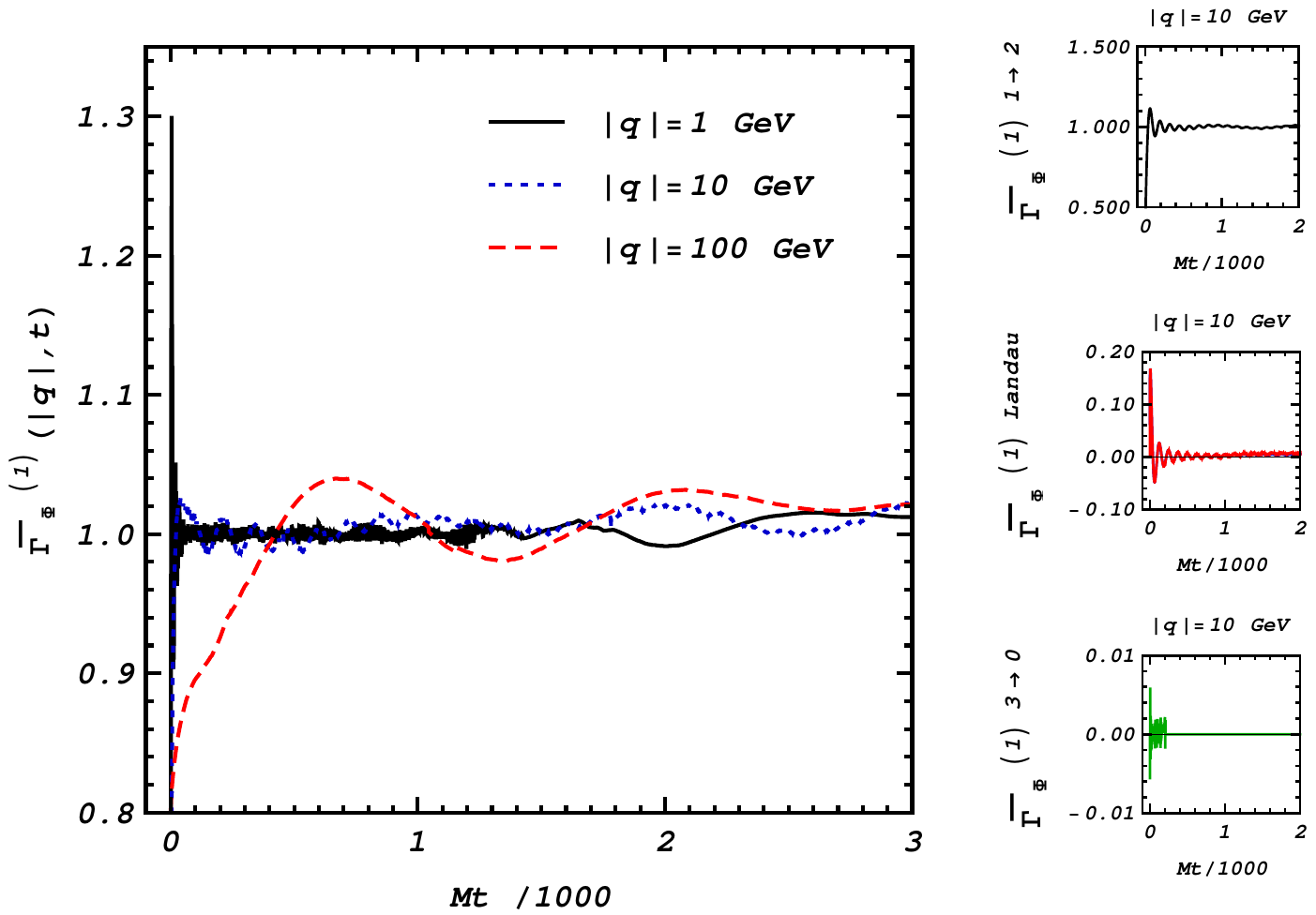}
    \vspace{-1em}
  \end{center}
  \caption{\emph{Left}: the total ratio $\bar{\Gamma}_{\Phi}^{(1)}$
    versus $Mt$, for on-shell decays with
    $|\mathbf{q}|\:=\:1\ \mathrm{GeV}$ (solid black),
    $10\ \mathrm{GeV}$ (blue dotted) and $100\ \mathrm{GeV}$ (red
    dashed). \emph{Right}: the separate contributions to
    $\bar{\Gamma}_{\Phi}^{(1)}$ for
    $|\mathbf{q}|\:=\:10\ \mathrm{GeV}$. The two Landau--damping
    contributions are equal up to numerical errors. }
  \label{fig:twidth}
  \vspace{-1em}
\end{figure}

In      figure~\ref{fig:twidth},      we      plot      the      ratio
$\bar{\Gamma}_{\Phi}^{(1)}(|\mathbf{q}|,t)\:                        =\:
\Gamma_{\Phi}^{(1)}(|\mathbf{q}|,t)/
\Gamma_{\Phi}^{(1)}(|\mathbf{q}|,t\:\to\:\infty)$        of        the
time-dependent width to its  late-time equilibrium value as a function
of  $Mt$  for  $q^2\:=\:M^2$.   In  addition,  we  show  the  separate
contributions of the four  processes in figure \ref{fig:kins}. We note
that the  oscillations in  the width have  time-dependent frequencies.
This non-Markovian  behaviour is inherent  to truly out-of-equilibrium
systems exhibiting so-called memory effects.

\section{Master time evolution equations for particle number densities}
\label{sec:evo}  

In  order   to  count   both  on-shell  and   off-shell  contributions
systematically, we `measure' the number of charges, rather than quanta
of energy. This avoids any need to identify `single-particle' energies
by a  quasi-particle approximation. We  begin by relating  the Noether
charge
\begin{equation}
  \label{eq:Q}
  \mathcal{Q}(x_0;\tilde{t}_i)\ =\ i\!\int\!\D{3}{\mathbf{x}}\;
  \Big(\,\Phi_{\mathrm{H}}^{\dag}(x;\tilde{t}_i)
  \pi_{\mathrm{H}}^{\dag}(x;\tilde{t}_i)
  \:-\:\pi_{\mathrm{H}}(x;\tilde{t}_i)
  \Phi_{\mathrm{H}}(x;\tilde{t}_i)\,\Big)
\end{equation}
to a charge density operator
$\mathcal{Q}(\mathbf{q},\mathbf{X},X_0;\tilde{t}_i)$ via
\begin{equation}
  \label{eq:Qdef}
  Q(X_0;\tilde{t}_i)\ =\ \!\int\D{3}{\mathbf{X}}\int\!\!
  \frac{\D{3}{\mathbf{q}}}{(2\pi)^3}\;
  \mathcal{Q}(\mathbf{q},\mathbf{X},X_0;\tilde{t}_i)\;.
\end{equation}
By         taking         the         equal-time        EEV         of
$\mathcal{Q}(\mathbf{q},\mathbf{X},X_0;\tilde{t}_i)$   and  extracting
the positive and negative  frequency particle components, we introduce
the following  definition of the  particle number density in  terms of
off-shell Green's functions:
\begin{equation}
  n(\mathbf{q},\mathbf{X},t)\ =\ \lim_{X_0\:\to\: t}\,2\!\int\!
  \frac{\D{}{q_0}}{2\pi}\!\int\!\!\frac{\D{4}{Q}}{(2\pi)^4}\;
  e^{-iQ\cdot X}\,\theta(q_0)q_0
  i\Delta_<(q+\tfrac{Q}{2},q-\tfrac{Q}{2},t;0)\;,
\end{equation}
where we have used the translational invariance of the CTP contour.

By partially inverting the CTP Schwinger--Dyson equation in
(\ref{eq:SD}), we derive the following master time evolution equation
for the statistical distribution function $f(\mathbf{q} +
\tfrac{\mathbf{Q}}{2},\mathbf{q}-\tfrac{\mathbf{Q}}{2},t)$:
\begin{align}
  \label{eq:evo}
  &\partial_{t}
  f(\mathbf{q}+\tfrac{\mathbf{Q}}{2},\mathbf{q}-\tfrac{\mathbf{Q}}{2},t)\:-\:
  2\!\iint\!\frac{\D{}{q_0}}{2\pi}\,\frac{\D{}{Q_0}}{2\pi}\;e^{-iQ_0t}
  \,\mathbf{q}\cdot\mathbf{Q}\,\theta(q_0)
  \Delta_<(q+\tfrac{Q}{2},q-\tfrac{Q}{2},t;0)\\&
  \qquad +\: \!\iint\!\frac{\D{}{q_0}}{2\pi}\,
  \frac{\D{}{Q_0}}{2\pi}\;e^{-iQ_0t}\,\theta(q_0)
  \Big(\,\mathscr{F}(q+\tfrac{Q}{2},q-\tfrac{Q}{2},t;0)\:+\:
  \mathscr{F}^{*}(q-\tfrac{Q}{2},q+\tfrac{Q}{2},t;0)\,\Big)\nonumber\\&
  \qquad \qquad =\ \!\iint\!\frac{\D{}{q_0}}{2\pi}\,
  \frac{\D{}{Q_0}}{2\pi}\;e^{-iQ_0t}\,\theta(q_0)
  \Big(\,\mathscr{C}(q+\tfrac{Q}{2},q-\tfrac{Q}{2},t;0)\:+\:
  \mathscr{C}^{*}(q-\tfrac{Q}{2},q+\tfrac{Q}{2},t;0)\,\Big)\;,\nonumber
\end{align}
where we have introduced
\begin{subequations}
\begin{align}
  \label{eq:fdef}
  \mathscr{F}(q+\tfrac{Q}{2},q-\tfrac{Q}{2},t;0)\ &\equiv\ 
  -\!\int\!\!\frac{\D{4}{k}}{(2\pi)^4}\;
  i\Pi_{\mathcal{P}}(q+\tfrac{Q}{2},k,t;0)\:
  i\Delta_<(k,q-\tfrac{Q}{2},t;0)\;,\\
  \label{eq:cdef}
  \mathscr{C}(q+\tfrac{Q}{2},q-\tfrac{Q}{2},t;0)\ &\equiv\
  \frac{1}{2}\!\int\!\!\frac{\D{4}{k}}{(2\pi)^4}\;
  \Big[\,i\Pi_>(q+\tfrac{Q}{2},k,t;0)\:
  i\Delta_<(k,q-\tfrac{Q}{2},t;0)
  \nonumber\\&\hspace{-7em}
  -\: i\Pi_<(q+\tfrac{Q}{2},k,t;0)\,
  \Big(\,i\Delta_>(k,q-\tfrac{Q}{2},t;0)
  \:-\:2i\Delta_{\mathcal{P}}(k,q-\tfrac{Q}{2},t;0)\,\Big)\Big]\;.
\end{align}
\end{subequations}
It  is  important  to  stress  here  that  (\ref{eq:evo})  provides  a
self-consistent  time evolution  equation for~$f$  valid  \emph{to all
  orders} in perturbation theory  and to \emph{all orders} in gradient
expansion. The  terms on the  LHS of (\ref{eq:evo}) may  be associated
with     the      total     derivative     in      the     phase-space
$(\mathbf{X},\ \mathbf{p})$, which  appears in the classical Boltzmann
transport   equation.   The   $\mathscr{F}$  terms   on  the   LHS  of
(\ref{eq:evo}) are the \emph{force}  terms, generated by the potential
due to  the dispersive part  of the self-energy and  the $\mathscr{C}$
terms n the RHS of~(\ref{eq:evo}) are the \emph{collision} terms.

Truncating  (\ref{eq:evo}) to leading  order in  a loopwise  sense, we
obtain  for our  simple scalar  theory, the  following  time evolution
equation for the $\Phi$ statistical distribution function:
\begin{align}
\label{eq:phievo}
\partial_tf_{\Phi}(|\mathbf{q}|,t)\ &=
\ -\frac{g^2}{2}\sum_{\alpha,\,\alpha_1,\,\alpha_2}\int\!\!\frac{\D{3}{\mathbf{k}}}{(2\pi)^3}\frac{1}{2E_{\Phi}(\mathbf{q})}\frac{1}{2E_{\chi}(\mathbf{k})}\frac{1}{2E_{\chi}(\mathbf{q}-\mathbf{k})}\nonumber
\\&\qquad \times \:
\frac{t}{2\pi}\,\mathrm{sinc}\Big[\Big(\alpha
    E_{\Phi}(\mathbf{q})-\alpha_1E_{\chi}(\mathbf{k})-\alpha_2E_{\chi}(\mathbf{q}-\mathbf{k})\Big)t/2\Big]\nonumber\\&\qquad
\times\:
\Big\{\pi+2\mathrm{Si}\Big[\Big(\alpha
    E_{\Phi}(\mathbf{q})+\alpha_1 E_{\chi}(\mathbf{k})+\alpha_2
    E_{\chi}(\mathbf{q}-\mathbf{k})\Big)t/2\Big]\Big\}\nonumber\\&\qquad
\times\:
\big\{\big[\theta(-\alpha)+f_{\Phi}(|\mathbf{q}|,t)\big]\big[\theta(\alpha_1)\big(1+f_{\chi}(|\mathbf{k}|,t)\big)+\theta(-\alpha_1)f_{\chi}^C(|\mathbf{k}|,t)\big]\nonumber\\&\quad
\quad \quad \quad \times \:
\big[\theta(\alpha_2)\big(1+f^C_{\chi}(|\mathbf{q}-\mathbf{k}|,t)\big)+\theta(-\alpha_2)f_{\chi}(|\mathbf{q}-\mathbf{k}|,t)\big]\nonumber\\&\qquad
- \:
\big[\theta(\alpha)+f_{\Phi}(|\mathbf{q}|,t)\big]\big[\theta(\alpha_1)f_{\chi}(|\mathbf{k}|,t)+\theta(-\alpha_1)\big(1+f_{\chi}^C(|\mathbf{k}|,t)\big)\big]\nonumber\\&\quad
\quad \quad \quad \times \:
\big[\theta(\alpha_2)f^C_{\chi}(|\mathbf{q}-\mathbf{k}|,t)+\theta(-\alpha_2)\big(1+f_{\chi}(|\mathbf{q}-\mathbf{k}|,t)\big)\big]\big\},
\end{align}
where  $\alpha,\ \alpha_1,\  \alpha_2\: =  \: \pm  1$. The  second and
third lines  of (\ref{eq:phievo})  encode the early-time  violation of
energy  conservation. This  leads  to the  non-Markovian evolution  of
memory   effects   and   evanescent   contributions   from   otherwise
kinematically-disallowed  processes.  Replacing   these  lines  by  an
energy-conserving  delta  function,   we  recover  the  semi-classical
Boltzmann transport  equation. However, given  the equilibrium initial
conditions  of   our  model,  the  artificial   imposition  of  energy
conservation   along  with  the   properties  of   the  Bose--Einstein
distribution ensure that the RHS  of (\ref{eq:phievo}) is zero for all
times. Thus, the semi-classical Boltzmann equation cannot describe the
re-thermalization of our simple model.  This is true also for gradient
expansions of Kadanoff--Baym equations  when truncated to zeroth order
in time-derivatives.  Hence, it is only  when energy-violating effects
are  systematically included,  as in  this new  perturbative approach,
that the dynamics of this re-thermalization are captured.

\section{Conclusions}
\label{sec:conc}

We have  obtained master time evolution equations  for particle number
densities that are  valid to all orders in  perturbation theory and to
all orders  in gradient expansion.   The loopwise truncation  of these
time  evolution equations  remains  valid to  all  orders in  gradient
expansion, capturing  the evolution on all time  scales, including the
transient   dynamics.    This  prompt   behaviour   is  dominated   by
energy-violating  processes that  lead to  non-Markovian  evolution of
memory  effects.  The  underlying perturbation  series are  built from
non-homogeneous   free  propagators   and   explicitly  time-dependent
vertices.   Due to  the systematic  treatment of  finite  boundary and
observation  times, these  diagrammatic  series remain  free of  pinch
singularities.

\ack

The   work   of   PM   and   AP   is  supported   in   part   by   the
Lancaster--Manchester--Sheffield  Consortium  for Fundamental  Physics
under STFC  grant ST/J000418/1.  AP also  acknowledges partial support
by an IPPP associateship from Durham University.

\section*{References}

\bibliographystyle{iopart-num} \bibliography{pTFTproc}

\end{document}